\documentclass[runningheads]{llncs}
\usepackage{graphicx}
\usepackage{svg}
\usepackage[labelfont=bf]{caption}
\usepackage{subcaption}
\usepackage{algorithm}
\usepackage{algorithmic}
\usepackage{amsmath}
\usepackage{array}
\usepackage{enumerate}
\usepackage{tabu}
\usepackage{booktabs}
\usepackage{tabularx}
\usepackage{ragged2e}
\usepackage{etoolbox}
\usepackage{siunitx}
\usepackage{multirow}

\begin{document}
\robustify\bfseries
\title{Rethinking Ultrasound Augmentation: \\ A Physics-Inspired Approach}
%
%
\author{Maria Tirindelli\textsuperscript{*,1}, Christine Eilers\textsuperscript{*,1}, Walter Simson\textsuperscript{1}, Magdalini Paschali\textsuperscript{1}, Mohammad Farid Azampour\textsuperscript{1,2}, Nassir Navab\textsuperscript{1,3}}
\authorrunning{Tirindelli et al.}
%
\institute{
\textsuperscript{1} Computer Aided Medical Procedures, Technische Universität M\"unchen, Germany \\
\textsuperscript{2} Sharif University of Technology,Tehran,Iran \\
\textsuperscript{3} Computer Aided Medical Procedures, John Hopkins University, Baltimore, MD, USA \\
\textsuperscript{*} These authors contributed equally to this work.}

%
\maketitle              
\begin{abstract}

Medical Ultrasound (US), despite its wide use, is characterized by artefacts and operator dependency. Those attributes hinder the gathering and utilization of US datasets for the training of Deep Neural Networks used for Computer-Assisted Intervention Systems. Data augmentation is commonly used to enhance model generalization and performance. However, common data augmentation techniques, such as affine transformations do not align with the physics of US and, when used carelessly can lead to unrealistic US images. To this end, we propose a set of physics-inspired transformations, including deformation, reverb and Signal-to-Noise Ratio, that we apply on US B-mode images for data augmentation. We evaluate our method on a new spine US dataset for the tasks of bone segmentation and classification. 

\end{abstract}
\section{Introduction}
\label{sec:intro}
\begin{figure}
    \centering
    \includegraphics[width=\textwidth]{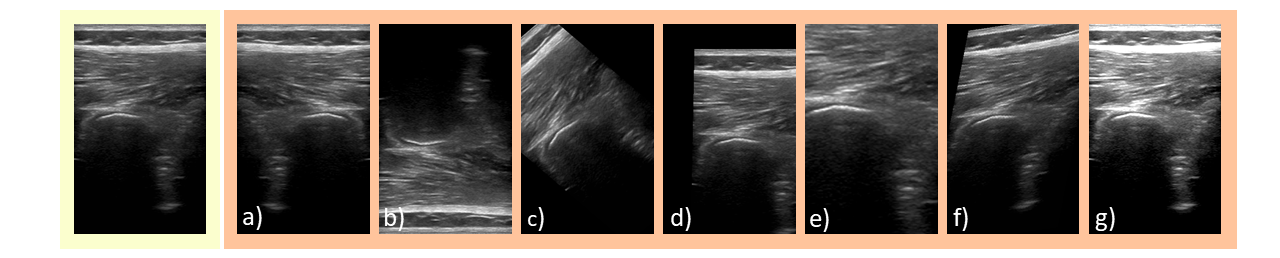}
    \caption{Classical augmentations applied to an US image: a) horizontal flip, b) vertical flip, c) rotation, d) translation, e) scaling, f) shearing, g) brightness adaption.}
    \label{fig:classical}
\end{figure}

Ultrasound (US) is a radiation-free, low-cost, and flexible imaging technique. However, US is characterized by low signal-to-noise ratio (SNR), artifacts, such as phase aberration, reverberation or scattering and an increased operator dependency. These attributes account for high variability in US images. Due to these shortcomings, interpreting US scans is an especially challenging task. \\
\noindent
With the rise of machine learning in medical imaging, Deep Neural Networks (DNN) are applied to US imaging to facilitate their interpretation~\cite{VanSloun}. Applications for these methods include bone segmentation and bone registration \cite{Wang} \cite{alsinan} \cite{Wangb} \cite{Hetherington}. Successful execution of these tasks can enable the development of innovative applications with an increased degree of autonomy, including robotic scans and diagnostic support \cite{Tirindelli} \cite{Esteban} \cite{Hase}. However, training DNNs requires large and diverse datasets. For medical US such datasets are limited due to the strenuous acquisition and annotation process. \\
\noindent
Moreover, US datasets that can be used for training DNNs should cover a wide range of variability, induced by operator dependency, US machines, acquisition parameters, and subject-anatomies. Such diverse US datasets could ensure the generalization of DNNs to an unseen test subject during a computer-aided intervention. \\
\noindent
In many computer vision tasks, data augmentation is applied to overcome the lack of training data~\cite{Shorten}. For DNNs trained with natural images, datasets are artificially increased by applying different transformation techniques. Augmentation methods could broadly be divided into two categories, namely synthetic data generation and image modification. In the first category, generative adversarial networks (GANs)~\cite{Goodfellow} are used to create synthetic data based on the limited available dataset. Such models have been applied by Zaman et al.~\cite{Zaman} on US images for different time gain compensations, different depths and different roll and pitch motions of the US probe. Synthetic augmentations, however, require a rather large amount of data to train the GAN and are, therefore, not suited for every scenario.\\
\noindent
In the second category, classical image modifications, such as random scaling, translations, rotations, and Gaussian noise additions, are applied to a dataset. In US this approach has been used in different works. Baka et al.~\cite{Baka} performed data augmentation through mirroring and small free-form deformations in US spine images for bone segmentation. Other works applied random shifting (\cite{Duong},~\cite{Nguyen},~\cite{Ungi}), random flipping (\cite{Duong},~\cite{Benjdira},~\cite{Hohlmann},~\cite{Qi},~\cite{Nguyen},~\cite{Luan},~\cite{Ungi}), different translations (\cite{Qi},~\cite{Patel}), rotation transformations (\cite{Benjdira},~\cite{Alsinan},~\cite{Patel},~\cite{Luan},~\cite{Ungi}) and varying brightness (\cite{Hohlmann},~\cite{Benjdira}) as an approach for data augmentation.\\
\noindent
However, these classical augmentations are based on the mechanisms behind optical cameras which strongly differ from the principles of US. Applying these transformations to US scans could create unrealistic US images inconsistent with the real variability in US acquisitions. A horizontal flipping (Fig.~\ref{fig:classical} a)) of the US image is still consistent with a 180-degree rotation of the US transducer. A vertical flipping, however, would result in a shadowing region between a strong reflector and the transducer, as shown in Fig.~\ref{fig:classical} b) which does not reflect the physical model of attenuation. Rotations and translations of the image (shown in~Fig. \ref{fig:classical} c), d)) result in displacement between the location of the transducer at the top of the image and the image content, therefore creating a gap between the wave source and reflective tissue.\\
\noindent
To this end, we propose a novel augmentation technique for US imaging that is inspired by the physical behavior of US. In this paper our contributions are: 
\begin{itemize}
    \item[$\circ$] We propose a novel data augmentation method, using a set of US image modifications that take into account realistic sources of variability in US. Our method can expand the size of US datasets and provide anatomically and physically consistent variability.
    \item[$\circ$] We comprehensively evaluate our method on the challenging tasks of bone segmentation and classification regarding the occurrence of bone per frame to showcase its suitability for a variety of crucial tasks for computer-assisted interventions.
    \item[$\circ$] We provide a new spine US dataset, that can be used for tasks such as automated robotic US scanning and intelligent localization of regions of interest. 
\end{itemize}

\begin{figure}[t]
   \centering
   \includegraphics[width=\textwidth]{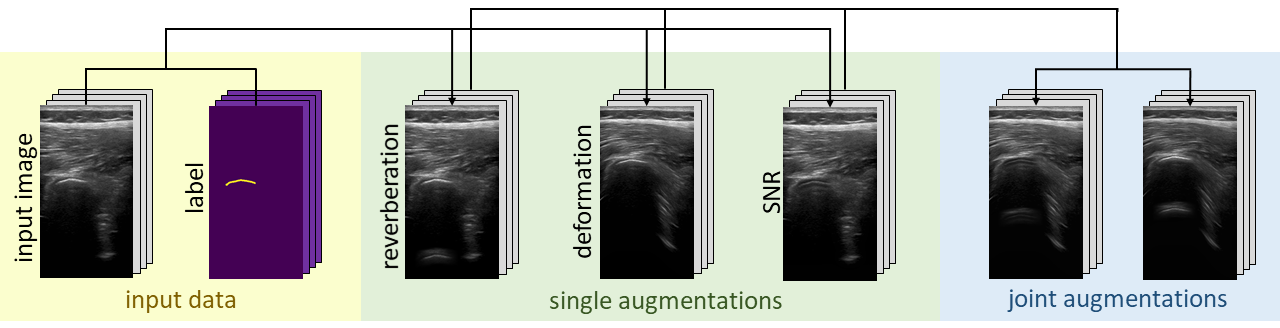}
    \caption{Our proposed data augmentation pipeline: from US images and corresponding bone masks (yellow), different augmentations are generated (green) and can be merged in the same image (blue).}
    \label{fig:header}
\end{figure}

\section{Methods}
\label{sec:methods}
The proposed augmentations are inspired by the physical model of wave propagation in medical US. These augmentations use the identification of regions of high-echogenicity or attenuation in order to model and augment the wave propagation and resulting B-mode image. We propose augmenting training data based on deformation due to probe pressure, reverberation artifacts and signal-to-noise-ratio, which have been tailored to linear US imaging. 
\begin{figure}[t]
    \centering
    \includegraphics[width=0.85\textwidth]{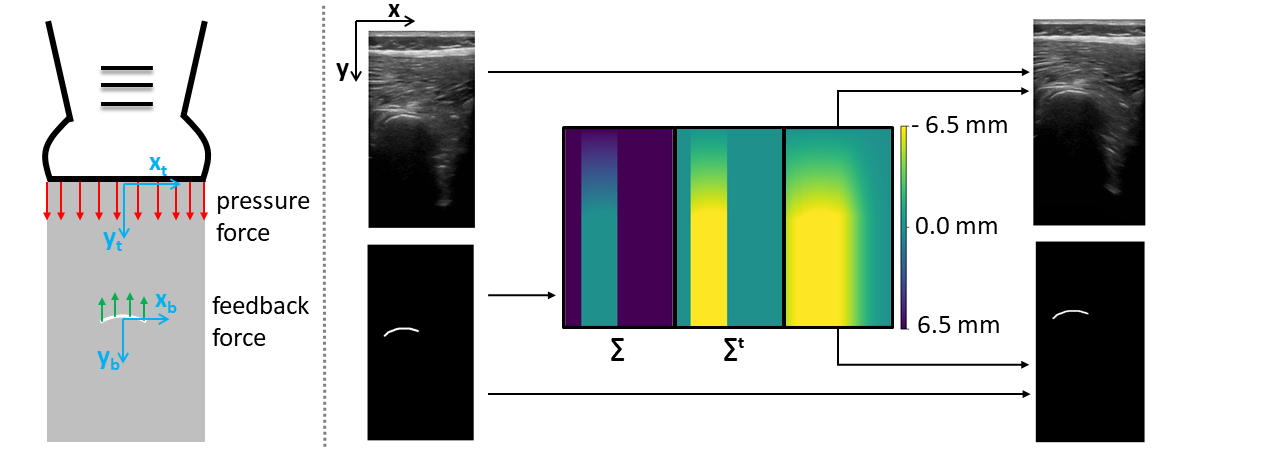}
    \caption{Deformation Pipeline: Left: The forces applied by the US probe (red arrows) and bones (green arrows), and the coordinate systems centered in the bone ($x_b, y_b$) and the probe centroids ($x_p, y_p$), respectively. Right: The original B-mode image and label. The label is used to compute the DF ($\Sigma$) induced by the probe pressure. To avoid unrealistic gap above the skin layer, the DF is expressed in the probe coordinate system ($\Sigma^t$) and blurred. $\Sigma^t$ is finally applied to both image and label.}
    \label{fig:deformation}
\end{figure}

\subsection{Deformation}
In this section, we propose a deformation model, seen in Fig.~\ref{fig:deformation} that simulates deformations from variations in the contact pressure of the US probe on interrogated tissues. 
Assuming a transducer displacement $d_{\text{probe}}$ for a given augmentation, we derive its respective Displacement Field (DF) for the deformed medium. This DF is then applied to the input B-mode image resulting in our proposed augmentation.
The DF is generated from a simplified physical model of the interactions of soft tissues, bones and the transducer. Specifically, we assume that soft-tissues are isotropic and homogeneous, that the probe pressure is applied along the axial direction and that the bone is a static body without deformation or transformation. Furthermore, we assume that only tissues between the probe and bone undergo compression and lateral tissue deformation is negligible.
\noindent
In lateral tissues there is no deformation, thus the DF is constant and equal to $d_{\text{probe}}$. In tissues laying between bone and probe, we can derive the deformation components as 
$\epsilon _{yy} = F/EA$, where $A$ is the transducer area and $F$ is the force applied by the probe and $E$ is the Young's Modulus.  This can be rewritten as $F = d_{\text{probe}}EA/y_{\text{probe}}$, where $y_{\text{probe}}$ is the position of the probe in the bone coordinate system. Hence, we have:

\begin{equation}
\Sigma(x, y) = \int \epsilon _{yy} dy \bigg|_{\Sigma(x, 0)=0} = -H(-y)\dfrac{F}{EA} y = -H(-y) \dfrac{d_\text{probe}}{y_\text{probe}} y
\end{equation}
where $\Sigma(x, y)$ is the DF. \\
\noindent
In order to avoid unrealistic gaps between skin and transducer, we rewrite the displacements in the transducer coordinate system. In this coordinate system $y_{\text{bone}}$ is the position of the bone with respect to the probe.
This corresponds to applying an offset of $d_{\text{probe}}$ to the DF equation.  
To ensure smooth transitions between regions with large and small displacements, we add lateral gaussian smoothing. The transducer translation  $d_{\text{probe}}$ is randomly sampled in our US augmentation. Pseudo-code is shown in Algorithm 1.
\begin{algorithm}
\caption{Computation of $\Sigma^t$ given $d_{\text{probe}}$}
\begin{algorithmic} 
\FOR{ $i, j = 1:\text{Width, Height}$}

\STATE $\Sigma(i, j)^t = - d_{\text{probe}}$ \textbf{if} $(i, j)$ is bone or below bone
\STATE $\Sigma(i, j)^t = -d_{\text{probe}} / y_\text{bone} \cdot j $ \textbf{if} $(i, j)$ is above bone
\STATE $\Sigma(i, j)^t = 0$ \textbf{else}
\ENDFOR
\STATE smooth($\Sigma^t (i, j)$)
\end{algorithmic}
\end{algorithm}

\subsection{Reverberation}

\begin{figure}[t]
    \centering
    \includegraphics[scale=0.3]{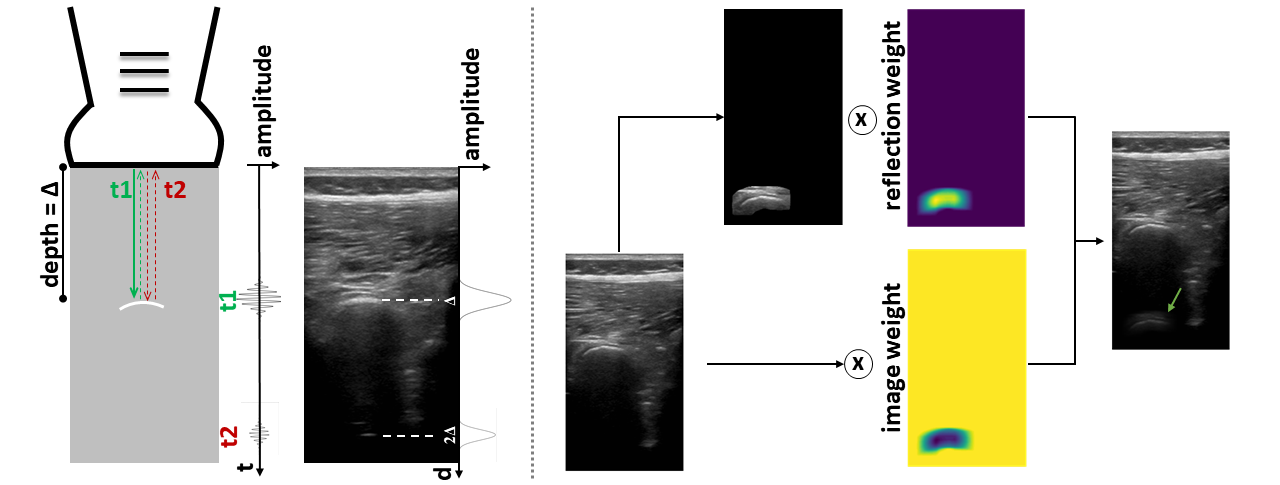}
    \caption{Reverberation Pipeline. Left: The transmitted signal and the echos generated by a bone at depth $\Delta$. In the time-amplitude plot the signals generated by the first echo (green arrow) and reverberation echo (red arrow) are reported. The first echo generates a bright area in the B-mode image at $\Delta$, while the reverberation echo generates an artifact at $2\Delta$. Right: The generation of reverberation artifacts on B-mode data. The patch containing the bone is extracted and replicated at multiples of $\Delta$. The replicated patch is weighted and summed up to the original image.}
    \label{fig:multiple_reflection}
\end{figure}

In this Section we propose a method to simulate reverberation artifacts on US data as can be seen in Fig.~\ref{fig:multiple_reflection}. Our model uses a ray-based approximation and assumes constant a speed of sound $c$ in tissues. Following these assumptions, a highly reflective structure at depth $\Delta$ in the interrogated tissue will generate an echo at $\Delta = \dfrac{c t_1}{2}$. If the echo intensity is sufficiently high, the signal is again reflected at the tissue-transducer interface back once more at the tissue-interface level. This generates an additional echo in the recorded signal, located at $\Delta_r$:
\begin{equation}
\Delta_{r} = 2\Delta = t_{1}c = \dfrac{t_{2}c}{2}
\end{equation}
\noindent
Multiple reflections, commonly known as reverberation, can occur several times, generating reverberation artifacts located at multiples of $\Delta$ in the B-mode image. 

\noindent
In the proposed method we simulate reverberation artifacts, by first computing the bone centroid $(x_c, y_c)$ on the label image. Secondly,  we shift the image patch containing the bone along the axially by $2y_c$. Finally, we compute the resulting image as the weighted sum of the shifted patch and the original image. The multiplication weights for the reverberation region $w_r$ are defined as the shifted label filtered with a Gaussian filter (kernel size=45 px, sigma=20 px) and scaled by a factor $r_i$, which we call reverberation intensity. The multiplication weights for the original image are defined as $1 - w_r$.
The factor $r_i$ defines the intensity of the reverberation artifact and is randomly sampled during the augmentation.

\subsection{Signal-to-Noise Ratio}
\begin{figure}[t]
    \centering
    \includegraphics[scale=0.4]{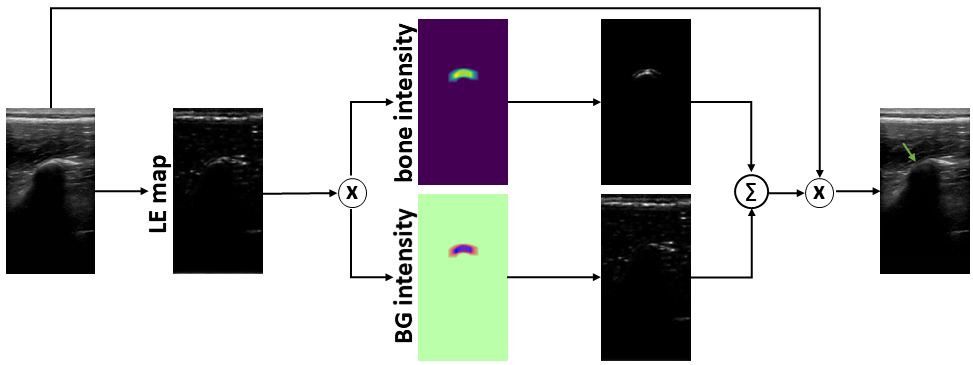}
    \caption{SNR augmentation pipeline. The local energy (LE) map is computed based on the input image. This image is then multiplied with the label or its inverse and the intensity values to obtain tuned local energy maps for bone and background (BG) regions. The overall tuned local energy is computed by adding the bone and background parts and the augmented image is obtained by multiplication of the input image with the tuned local energy.} 
    \label{fig:noise}
\end{figure}

Signal-to-Noise Ratio (SNR) provides information on the relation between the amount of desired signal and the background noise. For bone segmentation, the signal is given by coherent structures such as bone, and the noise consists of the background speckle. 
In our method we propose to tune the SNR in B-Mode data by first extracting and scaling the signal in the image to enhance or reduce the SNR, e.g. the visibility of bones compared to the background. The proposed pipeline is shown in Fig.~\ref{fig:noise}. To extract signal from the B-mode, we use the method proposed by \cite{Bridge} to compute local energy (LE) maps from B-Mode data. LE provides pixel-wise information on the signal energy at each location in space and is defined as:  
\begin{equation}
LE(x)  =  f_e(x)^2 + f_{o1}(x)^2 + f_{o2}(x)^2
\end{equation}
where $f_e(x)$, $f_{o1}(x)$ and $f_{o2}(x)$ are the components of the monogenic signal extracted from the US data. 
\noindent
To tune the SNR in the US data, we first normalize the images by the LE. Afterwards, we tune the LE maps by multiplying bone and non-bone pixels by different scaling factors, $i_{b}$ and $i_{bg}$. These factors define the scaling between signal and background energies. For $i_b > i_{bg}$ the bone structure is enhanced, while for $i_b < i_{bg}$ the background is intensified, making the bone less visible. Both factors are randomly sampled during the augmentation. Finally, we re-scale the B-mode image by the tuned LE map, thus obtaining a new image with an enhanced or reduced SNR.

\section{Experiments and results}
\label{sec:results}
We evaluated the proposed US augmentations on the tasks of bone segmentation and bone classification. 

\noindent
\textbf{Model architecture}: For the segmentation task we used a U-Net architecture~\cite{Ronneberger}, while for classification we use DenseNet121~\cite{Huang}. For training we used learning rate of 0.01, Adam optimizer and Binary Cross Entropy loss function.


\begin{table}[t]
\caption{Value ranges for the classical and proposed augmentation parameters selection during training. For each algorithm, the parameter is uniformly sampled from the value range.}\label{training_parameters}
\centering
\begin{tabular}{p{0.2\textwidth}p{0.5\textwidth}p{0.2\textwidth}}
\textbf{Augmentation} & \textbf{Parameter} & \textbf{Value Range} \\
\hline
\multirow{5}{*}{ \textbf{Classical} }  & rotation in degree & -10 - 10 \\
                           & translation (horizontal and vertical) & 0.2, 0.2 \\
      			   & scaling in both axis & 1, 1 \\
                          & shearing in both axis  & 1, 1  \\
                          & brightness & 0.2 \\
\hline
\multirow{3}{*}{ \textbf{Proposed}}  & deformation - $d_{\text{probe}}$ & 30 - 100 \\
		      & reverb - $r_i$ & 0.50 - 0.9  \\
                     & SNR - $i_b, i_{bg}$ & 0.70 - 1.40 \\
\end{tabular}
\end{table}

\noindent
\textbf{Dataset Acquisition}: Our models were trained on an US dataset consisting of 10.656 US images of 24 healthy volunteers with a BMI ranging from 20 to 25. The US scans were acquired with a Zonare z.one ultra sp Convertible Ultrasound System (Zonare Medical Systems Inc.) with a L8-3 linear probe. We set the image depth to 7cm, the gain to 92 \%, the frequency to 10 MHz and the sampling rate to 13 Hz.

\noindent
\textbf{Segmentation} For the bone segmentation task, we used 5.284 frames containing bones. On each of these frames, bones were annotated pixel-wise. The dataset was split subject-level to training (3972 frames from 17 subjects), validation (782 frames from 5 subjects) and testing (500 frames from 2 subjects).

\noindent
\textbf{Classification} For the bone classification task, we leveraged 5.692 US frames from 22 subjects. We split the dataset into two classes: one containing at least one bone per frame, the other containing no bone in each frame. The dataset was split subject-level to training (3821 frames from 16 subjects), validation (1037 from 4 subjects) and testing (834 frames from 2 subjects).

\noindent
\textbf{Implementation} Our method was implemented in PyTorch 1.7.1 and the DNNs were trained on a NVIDIA Titan V 12 GB HBM2 using Polyaxon\footnote{https://polyaxon.com/}. The source code for our method and our dataset will both be made public upon acceptance. For each task we evaluated the network performance using different augmentation techniques, that are shown in detail in Table~\ref{training_parameters}. Augmentations were only applied in training for both segmentation and classification tasks. 

\noindent
\textbf{Evaluation Metrics} All our models were trained using 5-fold cross validation. For the segmentation model we report the average Dice Score (DSC) and Hausdorff Distance (HDF) with standard deviation. Regarding classification, we evaluate average accuracy (ACC) and F1-score (F1) with standard deviation. 



\begin{table}[t]
    \centering
    \caption{Baseline comparisons for bone segmentation with UNet and bone classification with DenseNet. We report the average metrics over 5-fold cross validation along with their respective standard deviation ($\pm$).}
    \begin{tabu} to \textwidth { X[1.1c] X[c] X[c] X[c] X[c] }
         & \multicolumn{2}{c}{\textbf{Segmentation}} & \multicolumn{2}{c}{\textbf{Classification}} \\ 
        \cmidrule(){2-5}
          & \centering{DSC} & \centering{HDF} & \centering{ACC} & \centering{F1}  \\
        \cmidrule(lr){2-3} \cmidrule(l){4-5} 
\textbf{None} & 0.589 $\pm$ 0.07   & 20.72 $\pm$ 3.84 & 0.876 $\pm$ 0.06 & 0.770 $\pm$ 0.15 \\
\textbf{Classical} & 0.625 $\pm$ 0.03 & 17.76 $\pm$ 3.17 & \bfseries 0.883 $\pm$ 0.04 & 0.780 $\pm$ 0.09 \\
\textbf{Reverb} & 0.604 $\pm$ 0.03 & 19.71 $\pm$ 2.20 & \bfseries 0.883 $\pm$ 0.03 &  \bfseries 0.802 $\pm$ 0.04 \\
\textbf{Deformation} & \bfseries 0.626 $\pm$ 0.01 & 19.06 $\pm$ 3.63 & 0.865 $\pm$ 0.04 & 0.759 $\pm$ 0.11 \\
\textbf{SNR} & \bfseries 0.626 $\pm$ 0.02 & \bfseries 17.24 $\pm$ 1.83 & 0.877 $\pm$ 0.06 & 0.764 $\pm$ 0.16\\
\textbf{All} & 0.600 $\pm$ 0.02 &  17.32 $\pm$ 2.97 & 0.834 $\pm$ 0.02 & 0.742 $\pm$ 0.04
    \end{tabu}
    \label{tab:results}
\end{table}

\section{Results and Discussion}
Table~\ref{tab:results} showcases the results for the segmentation and classification networks for different augmentations. ``All'' denotes models trained with all three proposed augmentations. For the segmentation task, it can be seen that augmentation substantially improves the network performance both in terms of DSC by 2-4\% and HDF. This signifies the fact that data augmentation for the challenging task of bone segmentation is crucial and has a positive effect. The proposed transformations of Deformation and SNR outperform the classical augmentation by 1\% for DSC, while SNR achieves the lowest HDF for all techniques. Additionally, the models trained with the proposed transformations had the lowest standard deviation among folds, highlighting their consistency.  Combining all proposed transformations, the performance slightly drops by 2\%, which could be attributed to the fact that the resulting combination of transformations is not present on our unseen test set. \\
\noindent
For the task of bone classification, the model trained with Reverb augmentation outperforms the other methods in terms of F1-Score by 2-7\%, highlighting the potential of this transformation. Consistent with our segmentation findings, combining all transformations did not provide an additional benefit. Overall, the results indicated that Deformation and SNR transformations were beneficial for bone segmentation and Reverb for classification. 

\section{Conclusion}
\label{sec:conclusion}
In this conference paper we introduced a novel, physics-inspired augmentation method for US B-mode images. We proposed augmentations that tune the SNR, deformations and reverberations. Our augmentations were evaluated on a new US spine dataset for the tasks of bone segmentation and bone classification. The results showcased the impact of data augmentation on training DNNs for medical US. US augmentation with deformations and reverberation marginally outperformed the classical augmentation baseline, highlighting the potential of these transformation approaches. Future work includes investigating further realistic and anatomically-consistent US augmentations that contribute towards training generalizable CNNs that can be leveraged for robotic US computer-assisted interventions. 



\begin{thebibliography}{140}
\bibitem{VanSloun}
Van Sloun, R. J. G,, Cohen, R., Eldar, Y. C.: Deep learning in ultrasound imaging. In: Proceedings of the IEEE 108.1 (2019): 11-29.

\bibitem{Shorten}
Shorten, C., Khoshgoftaar, T. M.: A survey on image data augmentation for deep learning. In: Journal of Big Data 6.1 (2019): 1-48.

\bibitem{Goodfellow}
I. J., Goodfellow, et al. "Generative adversarial networks." arXiv preprint arXiv:1406.2661 (2014).

\bibitem{Zaman}
Zaman, A., Park, S.H., Bang, H., Park, C., Park, I., Joung, S.: Generative approach for data augmentation for deep learning-based bone surface segmentation from ultrasound images. In: Int J CARS (2020), vol. 15, pp. 931–941.
\doi{10.1007/s11548-020-02192-1}

\bibitem{Baka}
Baka, N., Leenstra, S., van Walsum, T.: Ultrasound Aided Vertebral Level Localization for Lumbar Surgery. In: IEEE Transactions on Medical Imaging, vol. 36, no. 10, pp. 2138-2147, Oct. 2017 \doi{10.1109/TMI.2017.2738612}.

\bibitem{Duong}
Duong, D. Q., Nguyen, K. -C. T., Kaipatur, N. R., Lou, E. H. M., Noga, M., Major, P. W., Le, L. H.: Fully Automated Segmentation of Alveolar Bone Using Deep Convolutional Neural Networks from Intraoral Ultrasound Images. In: 41st Annual International Conference of the IEEE Engineering in Medicine and Biology Society (EMBC), Berlin, Germany, 2019, pp. 6632-6635, \doi{10.1109/EMBC.2019.8857060}.

\bibitem{Hohlmann}
Hohlmann, B., Glanz, J., Radermacher, K.: Segmentation of the distal femur in ultrasound images. In: Current Directions in Biomedical Engineering 2020. 6(1): 20200034

\bibitem{Qi}
Qi, X., Voar, N., Riera, L., Sarangi, A., Youssef, G., Vives, M., Hacihaliloglu, I.: Automatic Scan Plane Identification from 2D Ultrasound for Pedicle Screw Guidance. In: CAOS 2018 (EPiC Series in Health Sciences, vol. 2), pp. 168–174

\bibitem{Benjdira}
Benjdira, B., Ouni, K., Al Rahhal, M. M., Albakr, A., Al-Habib, A., and Mahrous, E. (2020). Spinal cord segmentation in ultrasound medical imagery. In Applied Sciences, 10(4), 1370.

\bibitem{Alsinan}
Alsinan, A. Z., Vives, M., Patel, V., Hacihaliloglu, I.: Spine Surface Segmentation from Ultrasound Using Multi-feature Guided CNN. In: CAOS 2019 (EPiC Series in Health Sciences, vol. 3), pp. 6–10

\bibitem{Nguyen}
Nguyen, K. C. T., Duong, D. Q., Almeida, F. T., Major, P. W., Kaipatur, N. R., Pham, T. T., Lou, E. H. M., Noga, M., Punithakumar, K., and Le, L. H.: Alveolar Bone Segmentation in Intraoral Ultrasonographs with Machine Learning. In: Journal of Dental Research (2020), 99(9), 1054–1061. \doi{10.1177/0022034520920593}

\bibitem{Patel}
Patel, H., Hacihaliloglu, I.: Improved Automatic Bone Segmentation Using Large-Scale Simulated Ultrasound Data to Segment Real Ultrasound Bone Surface Data. In: IEEE 20th International Conference on Bioinformatics and Bioengineering (BIBE), Cincinnati, OH, USA, 2020, pp. 288-294, \doi{10.1109/BIBE50027.2020.00054}

\bibitem{Luan}
Luan, K., Li, Z., Li, J.: An efficient end-to-end CNN for segmentation of bone surfaces from ultrasound. In: Computerized Medical Imaging and Graphics, Volume 84, 2020, 101766, ISSN 0895-6111, \doi{10.1016/j.compmedimag.2020.101766}

\bibitem{Ungi} 
Ungi, T., Greer, H., Sunderland, K. R., Wu, V., Baum, Z.M.C., Schlenger, C., Oetgen, M., Cleary, K., Aylward, S. R., Fichtinger, G.: Automatic Spine Ultrasound Segmentation for Scoliosis Visualization and Measurement. In: in IEEE Transactions on Biomedical Engineering, vol. 67, no. 11, pp. 3234-3241, Nov. 2020, \doi{10.1109/TBME.2020.2980540}

\bibitem{Bridge}
Bridge, C. P., and Noble, J. A. (2015, April). Object localisation in fetal ultrasound images using invariant features. In 2015 IEEE 12th International Symposium on Biomedical Imaging (ISBI) (pp. 156-159). IEEE.

\bibitem{Ronneberger}
Ronneberger, O., Fischer, P., and Brox, T. (2015, October). U-net: Convolutional networks for biomedical image segmentation. In International Conference on Medical image computing and computer-assisted intervention (pp. 234-241). Springer, Cham.

\bibitem{Huang}
Huang, G., Liu, Z., Van Der Maaten, L., and Weinberger, K. Q. (2017). Densely connected convolutional networks. In Proceedings of the IEEE conference on computer vision and pattern recognition (pp. 4700-4708).


\bibitem{Tirindelli}
Tirindelli, M., Victorova, M., Esteban, J., Kim, S. T., Navarro-Alarcon, D., Zheng, Y. P., and Navab, N. (2020). Force-ultrasound fusion: Bringing spine robotic-us to the next “level”. IEEE Robotics and Automation Letters, 5(4), 5661-5668.

\bibitem{Esteban}
Esteban, J., Simson, W., Witzig, S. R., Rienm{\"u}ller, A., Virga, S., Frisch, B., ... and Hennersperger, C. (2018). Robotic ultrasound-guided facet joint insertion. International journal of computer assisted radiology and surgery, 13(6), 895-904.

\bibitem{Hase}
Hase, H., Azampour, M. F., Tirindelli, M., Paschali, M., Simson, W., Fatemizadeh, E., and Navab, N. (2020). Ultrasound-guided robotic navigation with deep reinforcement learning. In IEEE/RSJ International Conference on Intelligent Robots and Systems (IROS).

\bibitem{Wang}
Wang, P., Vives, M., Patel, V. M., and Hacihaliloglu, I. (2020). Robust real-time bone surfaces segmentation from ultrasound using a local phase tensor-guided CNN. International Journal of Computer Assisted Radiology and Surgery, 15, 1127-1135.

\bibitem{alsinan}
Alsinan, A., Vives, M., Patel, V., and Hacihaliloglu, I. (2019). Spine surface segmentation from ultrasound using multi-feature guided cnn. CAOS, 3, 6-10.

\bibitem{Wangb}
Wang, P., Patel, V. M., and Hacihaliloglu, I. (2018, September). Simultaneous segmentation and classification of bone surfaces from ultrasound using a multi-feature guided CNN. In International conference on medical image computing and computer-assisted intervention (pp. 134-142). Springer, Cham.

\bibitem{Hetherington}
Hetherington, J., Lessoway, V., Gunka, V., Abolmaesumi, P., and Rohling, R. (2017). SLIDE: automatic spine level identification system using a deep convolutional neural network. International journal of computer assisted radiology and surgery, 12(7), 1189-1198.

\end{thebibliography}
\end{document}